
\input amstex
\magnification 1200
\documentstyle{amsppt}
\NoBlackBoxes
\NoRunningHeads
\def\g{\frak g}

\def\Z{\Bbb Z}
\def\C{\Bbb C}

\def\d{\partial}

\def\Tr{\text{\rm Tr}}
\def\l{\lambda}

\def\Id{\text{Id}}
\topmatter
\title Center of a quantum affine algebra
       at the critical level
  \endtitle
\author {\rm {\bf Jintai Ding and Pavel Etingof} \linebreak
	\vskip .1in
   Department of Mathematics\linebreak
   Yale University\linebreak
   New Haven, CT 06520, USA\linebreak
   e-mail: ding\@math.yale.edu, etingof\@math.yale.edu}
\endauthor
\endtopmatter

\centerline{March 9, 1994}
\centerline{Submitted to Math. Research Letters}
\vskip .1in
\centerline{Dedicated to the memory of Ansgar Schnizer,}
\centerline{our friend and colleague who died tragically in Japan}
\vskip .1in
\centerline{\bf Abstract}
\vskip .1in

We construct central elements in a completion of the quantum affine
algebra $U_q(\hat\g)$ at the critical level $c=-g$
from the universal $R$-matrix ($g$ being the dual
Coxeter number of the simple Lie algebra $\g$), using the
method of Reshetikhin and Semenov-Tian-Shansky \cite{RS}.
This construction defines an action of the Grothendieck algebra
of the category of finite-dimensional representations of $U_q(\hat\g)$
on any $U_q(\hat\g)$-module from category $\Cal O$ with $c=-g$.
We explain the connection between the central elements
from \cite{RS} and transfer matrices in statistical mechanics.
In the quasiclassical approximation this connection was explained
in \cite{FFR}, and it was mentioned that one could generalize it
to the quantum case to get Bethe vectors for transfer matrices.
Using this connection, we prove that the central elements
from \cite{RS} (for all finite dimensional representations)
applied to the highest weight vector of a generic
Verma module at the critical level generate the whole space of
singular vectors in this module.
We also compute the first term of the
quasiclassical expansion of the central elements near $q=1$, and show
that it always gives the Sugawara current with a certain coefficient.

\vskip .1in
\centerline{\bf 1. Central elements and singular vectors}
\vskip .1in

Let $\g$ be a simple Lie algebra over $\C$ of rank $r$.
Let $U_q(\hat \g)$ be the corresponding
quantum affine algebra, and let $U_q(\tilde \g)$
be its extension by the scaling element $d$ (see \cite{Dr1}; however,
our notations will be as in \cite{FR}).
We assume that $q$ is a formal parameter but sometimes we will use
the specialization $q=1$.
Let $U_q(\tilde\frak n^{\pm})$ be
subalgebras of $U_q(\tilde\g)$
generated by the positive and negative root elements,
respectively. Let $\{a_i,i\ge 0\}$ be a homogeneous basis of $U_q(\tilde\frak
n^{+})$ ($a_0=1$), and let
$\{a^i\}$ be the dual basis of $U_q(\tilde\frak n^{-})$
(with respect to the Drinfeld pairing, \cite{Dr1}). Then the universal
$R$-matrix of $U_q(\tilde\frak g)$ is (see \cite{Dr1}):
$$
\tilde\Cal R=q^{c\otimes d+d\otimes c}\Cal R=
q^{c\otimes d+d\otimes c+\sum_{j=1}^r X_j\otimes
X_j}(1+\sum_{i>0}a_i\otimes a^i),\tag 1.1
$$
where $c$ is the central element, and
$X_j$ is an orthonormal basis of the Cartan subalgebra in $\g$
with respect to the invariant form $<,>$ normalized by
$<\theta,\theta>=2$ ($\theta$ is the maximal root of $\g$).

Let $V$ be a finite-dimensional representation of $U_q(\hat\g)$.
Let $\pi_V: U_q(\hat\g)\to \text{End}(V)$ be the corresponding
homomorphism. For $z\in\C^*$, let $V(z)$ be the representation
of $U_q(\hat\g)$ defined by $\pi_{V(z)}(a)=\pi_V(z^daz^{-d})$.

Consider the following quantum currents
(cf.\cite{FRT},\cite{RS},\cite{FR},\cite{DF}):
$$
L_V^+(z)=(\Id\otimes\pi_{V(z)})(\Cal R'),\
L_V^-(z)=(\Id\otimes\pi_{V(z)})(\Cal R^{-1}),\tag 1.2
$$
where $\Cal R'$ is obtained from $\Cal R$ by permutation of factors.
They are series in $z$ with coefficients in
 some completion of $U_q(\hat\g)\otimes
\text{End}(V)$.

Now, following \cite{RS}, introduce the $L$-matrix
$$
L_V(z)=(q^{-cd}\otimes 1)L_V^+(z)(q^{cd}\otimes 1)L_V^-(z)^{-1},\tag 1.3
$$
and consider the current
$$
l_V(z)=\Tr|_{V}\bigl((1\otimes q^{2\rho})L_V(z)\bigr).\tag 1.4
$$
This is a formal series in $z$ infinite in both directions, and
its components belong to a completion of the quantum affine algebra.
However:

\proclaim{Lemma 1.1} Let $U$ be any highest weight module over
$U_q(\hat\g)$. Then for any $u\in U$ $l_V(z)u\in U((z))$ (i.e. it
is a series in $z$ finite in the negative direction and its
coefficients belong to $U$).
\endproclaim

This lemma follows from the definition of the universal $R$-matrix.

It turns out that at the critical level $l_V(z)$ becomes central.
This construction of central elements is due to Reshetikhin and
Semenov-Tian-Shansky \cite{RS}, and is analogous to the construction
of the center for $U_q(\g)$ due to Drinfeld \cite{Dr2} and Reshetikhin
\cite{R}.

\proclaim{Theorem 1.2} (\cite{RS}) Let $U$
be any highest weight module over
$U_q(\hat\g)$ with central charge $c=-g$, where $g$ is the dual
Coxeter number of $\g$. Then $l_V(z)a=al_V(z)$ in $U$ for any
$a\in U_q(\hat\g)$.
\endproclaim

Let $l_V^0$ be the central element of $U_q(\g)$ defined
by $l_V^0=\Tr|_{V}(R_{21}R)$, where $R$ is the universal $R$-matrix
for $U_q(\g)$ (\cite{Dr2},\cite{R}).
 If $U$ is of highest weight $\lambda$ then $l_V^0$ is a scalar in $U$:
 $l_V^0|_U=\chi_V(q^{2(\lambda+\rho)})$, where $\chi_V$ is the character
of $V$ as a $U_q(\g)$-module.

\proclaim{Proposition 1.3} (properties of $l_V$) Let $U$ be as in
Theorem 1.2. Then in $U$:

(i) $l_V(z)$ is regular at $z=0$ and $l_V(0)=l_V^0$;

(ii) for any exact sequence
$0\to V_1\to V_2\to V_3\to 0$ of finite-dimensional
representations of $U_q(\hat\g)$ one has $l_{V_2}(z)=l_{V_1}(z)+l_{V_3}(z)$;

(iii)
$$
l_{V(z_1)}(z_2)=l_V(z_1z_2);\tag 1.5
$$

(iv)
$$
l_{V_1}(z_1)l_{V_2}(z_2)=l_{V_1(z_1/z_2)\otimes V_2}(z_2);\tag 1.6
$$
\endproclaim

\demo{Proof} (i) Let $u_0\in U$ be the highest weight vector.
 Let $l_V[n]$ denote the coefficient
to $z^{n}$ in $l_V(z)$. Then $l_V(z)u_0=l_V^0u_0+\sum_{n>0}z^{n}l_V[n]u_0$.
Let $u\in U$. Pick $a\in U_q(\hat\frak n^-)$ such that $u=au_0$. Then
by Theorem 1.2 $l_V(z)u=l_V^0u+\sum_{n>0}z^{n}l_V[n]u$.

(ii) The matrix $(1\otimes q^{2\rho})L_V(z)$ is
block-triangular, and its trace
 is the sum of the traces of its diagonal blocks.

(iii) Straightforward.

(iv) Using Theorem 1.2 and property (iii), we get
$$
\gather
l_{V_1}(z_1)l_{V_2}(z_2)=
\Tr|_{V_2}((1\otimes q^{2\rho})
L_{V_2}^+(q^gz)l_{V_1(z_1/z_2)}(z_2)L_{V_2}^-(z)^{-1})=\\
\Tr|_{V_1(z_1/z_2)}\Tr|_{V_2}
((1\otimes q^{2\rho}\otimes q^{2\rho})
L_{V_2}^+(q^gz)L_{V_1(z_1/z_2)}^+(q^gz)L_{V_1(z_1/z_2)}^-(z)^{-1}
L_{V_2}^-(z)^{-1})=\\
\Tr|_{V_1(z_1/z_2)\otimes V_2}
((1\otimes q^{2\rho})
L_{V_1(z_1/z_2)\otimes V_2}^+(q^gz)L_{V_1(z_1/z_2)\otimes
V_2}^-(z)^{-1})=\\
l_{V_1(z_1/z_2)\otimes V_2}(z_2).\tag 1.7
\endgather
$$
$\square$\enddemo

{\bf Remark. }
If $U$ is a module from category $\Cal O$ but not necessarily highest
weight then property (i) no longer holds, and all Fourier components
of $l_V(z)$ could be nontrivial operators.
\vskip .05in

Properties (ii)-(iv) imply:

\proclaim{Corollary 1.4} The map $V\to l_V(1)$ defines an action
of the Grothendieck algebra $Gr$ of finite dimensional
representations of the quantum affine algebra $U_q(\hat\g)$ on
 any completed highest weight module $U$ over this
algebra with $c=-g$.
\endproclaim

{\bf Remark.} By definition, finite-dimensional representations with
the same Jordan-H\"older series correspond to the same element
in the Grothendieck algebra.
\vskip .05in

Let now $U$ be as in Theorem 2, and let $u_0$ be the highest
weight vector in $U$. Let $l_V[n]$ denote the coefficient
to $z^{n}$ in $l_V(z)$. Then for every $V$ and $n$ $l_V[n]u$ is
a singular vector in $U$. A natural question is: do such vectors
span the space of singular vectors in $U$ if the highest weight is
generic? We will prove that answer is positive.

More precisely,
let $\omega_1,...,\omega_r$ be the fundamental weights of $\g$. Let
$V_1,...,V_r$ be
 deformations of the fundamental representations, i.e. irreducible
finite-dimensional representations of $U_q(\hat\g)$
such that the restrictions of $V_j$ to $U_q(\g)$
 are $V_j=L_{\omega_j}\oplus\sum_{\omega<\omega_j}c_{\omega}
L_{\omega}$, where $L_{\omega}$ is the irreducible $U_q(\g)$ module
with highest weight $\omega$ (cf. \cite{Dr3},\cite{Dr4},\cite{CP}).
These representations are the simplest in the case
$\g=\frak{sl}_{r+1}$: in this case $V_i=\Lambda_q^{i}\C^{r+1}$
are just the quantum exterior powers pulled back from
$U_q(\g)$ to $U_q(\hat\g)$ by the evaluation homomorphism
$p: U_q(\hat\g)\to U_q(\g)$ defined in \cite{J}.

 Let $y_{jn}=l_{V_j}[n]$. Let $A$ be the free
polynomial algebra in $y_{jn}$, $j=1,...,r$, $n>0$.
Let $U_0$ be the space of singular vectors in $U$.
Then $U_0$ is an $A$-module, via $(y,u)\to yu$.

 Let $\lambda$ be generic. This means that we regard $S=q^{2(\l+\rho)}$
as an indeterminate taking values in the Cartan subgroup
of the Lie group corresponding to $\g$.
Let $U$ be the Verma module
with highest weight $\l$ and central charge $-g$.

\proclaim{Theorem 1.5} (Main result) For $q\ne 1$
the space
 $U_0$ is a free $A$-module of rank 1 generated
by $u_0$.
\endproclaim

The proof of this theorem will be given in Section 3.

{\bf Remark. } This result is a quantum analogue of the
theorem of Feigin and Frenkel \cite{FF} who showed
that certain central elements in the completion of $U(\hat\g)/(c+g)$
applied to a highest weight vector in a generic Verma module
generate the space of all singular vectors in this module.

\proclaim{Corollary 1.6}
The space $U_0$ is a cyclic $Gr$-module generated by $u_0$.
\endproclaim

\vskip .1in
\centerline {\bf 2. Central elements, Bethe vectors, and transfer matrices}
\vskip .1in

In Section 1 we defined a representation of the Grothendieck algebra
$Gr$ of finite-dimensional $U_q(\hat\g)$-modules on a critical level
module. Representations of this algebra also occur in statistical
mechanics as transfer matrices. In this section we will connect these
two representations.

Let $W$ be a $U_q({\hat\g})$-module with central charge 0.
Fix $\beta\in\frak h^*$.
Define {\it transfer matrices} to be the following
operators in $W$:
$$
T_V(z)=\Tr|_V(R^{VW}(z)(q^{\beta}\otimes 1)), \tag 2.1
$$
where $R^{VW}(z)$ is the projection of the universal R-matrix
$\Cal R$ to $V(z)\otimes W$, and $V$ is a finite-dimensional
representation of $U_q(\hat\g)$. Obviously, transfer matrices preserve
weight, i.e. commute with the Cartan subalgebra in $U_q(\g)$.
So we can restrict their action to a weight subspace in $W$, say
$W[\mu]$, $\mu\in\frak h^*$.

It follows from the definition of the R-matrix that
the transfer matrices are pairwise commutative and satisfy
properties (ii)-(iv) of Proposition 1.3. Hence, similarly to $l_V(z)$,
they define a representation of $Gr$. Therefore, it is natural
to ask if there is any relation between $l_V$ and $T_V$.
It turns out there is a close connection between them.
To explain this connection, we need to introduce intertwiners.

Let $\l=(\beta+\mu-2\rho)/2$, $\nu=\l-\mu$.
Let $M_{\lambda,-g}$ denote the Verma module over
$U_q(\hat\g)$ with highest weight $\lambda$ and central charge $-g$.
Consider intertwining operators
$$
\Phi: M_{\l,-g}\to M_{\nu, -g}^c\otimes W,\tag 2.2
$$
where $M^c$ denotes the (complete) dual module to $M$ twisted by the
Cartan involution (the contragredient module).
 Such operators were defined and studied in
\cite{FR}. The following fact is known about them:

\proclaim{Lemma 2.1} The space of intertwiners (2.2) is isomorphic to
the weight space $W[\l-\nu]$; the isomorphism is given by
$\Phi\to <\Phi>$, where $<\Phi>=<v_{\nu},\Phi v_{\l}>\in W$
(here $<,>$ is the Shapovalov form, and $v_{\l}$, $v_{\nu}$ are the
vaccum vectors of $M_{\l,-g}$, $M^c_{\nu,-g}$).
\endproclaim

Lemma 2.1 allows us to connect the elements $l_V$ and $T_V$.

\proclaim{Proposition 2.2}
$$
<\Phi l_V(z)>=T_V(z)<\Phi>.\tag 2.3
$$
\endproclaim

\demo{Proof} For the proof we need the commutation relation
$$
\Phi L^+_V(zq^{-g})=R^{VW}(z)L^+_V(zq^{-g})\Phi,\tag 2.4
$$
which is a special case of (4.47) in \cite{FR}.
Applying this relation, we get
$$
\Phi L_V(z)=R^{VW}(z)L^+_V(zq^{-g})\Phi L_V^-(z)^{-1},
$$
or
$$
\gather
<\Phi L_V(z)>=R^{VW}(z)<L^+_V(zq^{-g})\Phi L_V^-(z)^{-1}>=
R^{VW}(z)<q^{\sum x_i\otimes x_i}\Phi q^{\sum x_i\otimes x_i}>=\\
R^{VW}(z)(q^{\l+\nu}\otimes 1)<\Phi>=
R^{VW}(z)(q^{\beta-2\rho}\otimes 1)<\Phi>.\tag 2.5
\endgather
$$
Now, multiplying the $V$-components of both sides of (2.5) by
$q^{2\rho}$ and taking the trace, and using the definitions of $l_V$
and $T_V$, we get (2.3). $\square$\enddemo

In statistical mechanics one is interested in finding Bethe vectors --
common eigenvectors of transfer matrices. Let us characterize Bethe
vectors in the language of intertwiners.

Let $\beta$ be generic. Then any singular vector in $M_{\l,-g}$
is of weight $\l$ with respect to $\frak h$, so the submodule generated
by this vector is isomorphic to the module $M_{\l,-g}$. This
makes legitimate the following definition:

{\bf Definition.} We say that an intertwiner $\Phi$ is a Bethe operator
if its restriction to every Verma submodule in $M_{\l,-g}$ is
proportional to $\Phi$.

Then we have an obvious proposition.

\proclaim{Proposition 2.3}
If $\Phi$ is a Bethe operator then $<\Phi>$ is a Bethe vector.
\endproclaim

\vskip .1in
\centerline {\bf 3. Proof of Theorem 1.5}
\vskip .1in

We will work with the Drinfeld (loop) realization of quantum affine algebras.
In this realization the algebra $U_q(\hat\g)$ is described as an
algebra generated by elements $x_{ij}$, $\xi_{ij}^{\pm}$, and
$c$ (central element), where $1\le i\le r$, and $j\in \Z$, satisfying
the relations listed in \cite{Dr4},\cite{KhT}. These elements
are quantum analogues of $h_i\otimes t^j$, $e_i\otimes t^j$,
$f_i\otimes t^j$, $c$, in the affine Lie algebra $\hat\g$.
We need to use only the quotient of $U_q(\hat\g)$ by the relation
$c=0$, which we denote by $U_q(L\g)$ (quantum loop algebra).
In this algebra, we define three subalgebras: $U^+, H, U^-$, generated
by $\xi_{ij}^+$, $x_{ij}$, $\xi_{ij}^-$, respectively.
About these algebras we only need to know that $H$ is abelian,
and $[H,U^+]\subset U^+$, $[H,U^-]\subset U^-$.

By a Drinfeld weight we mean an infinite set of numbers
$D=\{ d_{ij}\}$, $i=1,...,r$, $j\in\Z$.
We will only use weights with $d_{ij}=0$, $j<0$.
 To every Drinfeld weight
$D$ one can associate a one-dimensional module over
$HU^-$ in which $x_{ij}$ acts by $d_{ij}$ and $U^+$ acts trivially.
We will denote this module also by $D$. Set
$W(D)=\text{Ind}_{HU^-}^{U_q(L\g)}D$. The module $W$ has the following
property: its weights with respect to the Cartan
subalgebra $\frak h$ are $D_0=(d_{10},...,d_{r0})$ and higher, and the
$D_0$-weight subspace is one-dimensional. Let $w_0$ be the generator
of this subspace. Then for any
$\l,\nu\in\frak h^*$ such that $\l=\nu+D_0$ there exists a unique
 intertwining operator $\Phi$ of the form
(2.2) with $<\Phi>=w_0$, where $W=W(D)$.

Since the transfer matrices defined in Section 3
preserve weight, we have $T_V(z)w_0=t_V(z,D)w_0$, where $t_V(z,D)$
is some series with scalar coefficients.

\proclaim{Lemma 3.1} Let $\phi_{ij}$, $i=1,...,r$, $j\ge 1$, be
arbitrary numbers. Let $\phi_i(z)=
\sum_{n=1}^{\infty}\phi_{in}z^n$.
Then the Drinfeld lowest weight $D$ can be chosen in such
a way that $\beta+D_0=2(\l+\rho)$ and
$t_{V_i}(z,D)=t_{V_i}(0,D)+\phi_i(z)$, $i=1,...,r$, and
the components of $D$ are rational functions in $q$ and $S=q^{2(\l+\rho)}$.
\endproclaim

\demo{Proof} For the proof we need an explicit realization of the
universal $R$-matrix (1.1). Such a realization was provided
by Khoroshkin and Tolstoy \cite{KhT}:
\vskip .1in
{\bf Proposition.} (\cite{KhT}, Eq. (42))
The universal $R$-matrix can be represented in the form
$$
\Cal R=\Cal R^+\Cal R^0\Cal R^- q^{\sum_{j=1}^r X_j\otimes X_j}, \tag 3.1
$$
where $\Cal R^{\pm}\in U^{\pm}\otimes U^{\mp}$ are of total degree 0, and
$$
\Cal R^0=\text{exp}(\sum_{n>0}\sum_{i,j=1}^r c_{ij}^nx_{i,n}
\otimes x_{j,-n}),\tag
3.2
$$
and $c_{ij}^n$ is the inverse matrix to
$$
\frac{q^{n<\alpha_i,\alpha_j>}-q^{-n<\alpha_i,\alpha_j>}}
{n(q^{<\alpha_i,\alpha_i>}-q^{-<\alpha_i,\alpha_i>})(q^{<\alpha_j,\alpha_j>}-
q^{-<\alpha_j,\alpha_j>})},\tag 3.3
$$
$\alpha_i$ being the simple roots of $\g$.
\vskip .1in

Fix $\beta\in \frak h^*$. Set $D_0=2(\l+\rho)-\beta$.
Using definition (2.1) of the transfer matrix
and the above proposition, we compute $t_V(z,D)$:
$$
\gather
t_V(z,D)=\Tr|_{V(z)}(\Cal R^0q^{\sum_{j=1}^r X_j\otimes X_j}(q^{\beta}\otimes
1))|_{\C w_0}=
\\
\Tr|_{V(z)}(\text{exp}(\sum_{n>0}\sum_{i,j=1}^r c_{ij}^n
d_{j,-n}x_{in})
S).\tag 3.4
\endgather
$$
Note that the terms $\Cal R^+$ and $\Cal R^-$ drop out since we are computing
the action of $\Cal R$ on the lowest weight vector.

Let $b_{in}=\sum_{j=1}^rc_{ij}^nd_{j,-n}$. Since $d_{j,-n}$ can be
chosen arbitrarily, and the matrices $c_{ij}^n$ are invertible by the
definition, the numbers $b_{in}$ can also be arbitrary. In terms of
them, equation (3.4) takes the form
$$
t_V(z,D)=
\Tr|_{V(z)}(\text{exp}(\sum_{n>0}\sum_{i=1}^r b_{in}x_{in})S).\tag 3.5
$$

We must show that $t_{V_i}(z,D)$ can be made arbitrary Taylor series.
By deformation argument, it is enough to show this when  $q=1$.
Thus we must consider the quasiclassical limit of $V_i$.
This limit has the form $V_i|_{q=1}=L_{\omega_i}(1)\oplus \sum_j
M_{ij}$, where $M_{ij}$ are irreducible modules over $\hat\g$:
$M_{ij}=\otimes_m N_{ijm}(t_m)$, where $N_{ijm}$ are irreducible
$\g$-modules, and $N_{ijm}(t_m)$ are corresponding evaluation modules
with some parameters $t_m$. Also, $\lim_{q\to 1}x_{in}=h_i\otimes
t^n$, $h_i=h_{\alpha_i}$. This information allows us to compute the
right hand side of (3.5).

Let $\chi_i$ denote the characters of $L_{\omega_i}$. Let
$\psi_i(z)=\sum_{n>0}b_{in}z^n$. Note that
$\psi_i$ can be arbitrary Taylor series with zero free
term if the weight $D$ is suitably chosen.
We have
$$
t_{V_i}(z,D)|_{q=1}=\chi_i(e^{\sum_l\psi_l(z)h_l}S)+
\sum_j\prod_m \text{Ch}_{N_{ijm}}(e^{\sum_l\psi_l(zz_m)h_l}S),\tag 3.6
$$
where $\text{Ch}_{V}$ denotes the character of $V$ as a $\g$-module.

{\bf Remark. } If $\g=\frak{sl}_{r+1}$ then only the first term occurs
on the right hand side of (3.6).

Now, using (3.6), we can compute the coefficients $b_{jn}$ recursively
for any given series $t_{V_i}(z,D)$. At each step we will have to
solve a system of linear equations whose matrix is
$$
a_{il}^n=\Tr|_{L_{\omega_i}}(h_lS)+\sum_{j,m}z_m^n\Tr|_{N_{ijm}}(h_lS)
\text{Ch}_{N_{ijm}}(S)^{-1}\text{Ch}_{M_{ij}}(S), 1\le i,l\le n.\tag 3.7
$$
So we must show that the determinant of this matrix is not identically
zero for any $n>0$.

Using the fact that the highest weights of $N_{ijm}$
are lower than $\omega_i$, and the formula
$\Tr|_{V}(h_iS)=\frac{\d}{\d h_i}\Tr_{V}(S)$, we find that
only the first term in (3.7) contributes to the determinant, and
therefore this determinant is equal to the Jacobian of the
fundamental characters, i.e. the Weyl denominator:
$\text{det}(a_{il}^n)=\frac{\d(\chi_1,...,\chi_r)}{\d(h_1,...,h_r)}$,
which is not identically zero. This proves the lemma.
$\square$\enddemo

\proclaim{Lemma 3.2} For any $n>0$, the operators $l_{V_i}[j]$
in $U$ are algebraically independent for $i=1,...,r$, $j=1,...,n$.
\endproclaim

\demo{Proof} Let $t_V(z,D)=\sum_{n\ge 0}t_V(D)[n]z^n$.
Then Proposition 2.2 implies that
$<\Phi l_V[n]>=t_V(D)[n]<\Phi>$. Therefore, if there were a nontrivial
polynomial relation between the elements $l_{V_i}[j]$ in $U=M_{\l,-g}$, say
$P(\{l_{V_i}[j]\})=0$, the same polynomial relation would have to hold
for the numbers $t_{V_i}(D)[j]$ for any $D$. But this is impossible
since by Lemma 3.1 these numbers can be arbitrary.$\square$\enddemo

Now we can finish the proof of the theorem. Lemma 3.2 implies that
the action of the algebra $A$ on the vacuum vector $u_0$ defines
an embedding of $A$ into $U_0$ as a graded vector space. On the other
hand, the character of $U_0$ at $q=1$ is equal to the character of
$A$. This implies that for $q\ne 1$ the dimension of the homogeneous
subspace in $U_0$ of degree $d$ is at most that for $A$.
This means that the
embedding of $A$ into $U_0$
is an isomorphism, i.e. that $U_0$ is a free module of rank
1 over $A$. The theorem is proved. $\square$

\vskip .1in
\centerline{\bf 4. Quasiclassical limit}
\vskip .1in

In this section we compute the first term of the quasiclassical
expansion of the central elements introduced in Section 1.

Let the number $C_V$ be defined by $\Tr|_V(ab)=C_V<a,b>$, $a,b\in\g$
(here we abuse the notation by using the same symbol $V$ for the
quasiclassical limit of $V$ at $q\to 1$ regarded as a $\g$-module).

\proclaim{Theorem 4.1} In any $U_q(\hat\g)$-module $U$ from the
category $\Cal O$ with $c=-g$
$$
\lim_{q\to
1}\frac{l_V(z)-\text{dim}V}{(q-q^{-1})^2}=C_V(\sum_{j\in\Z}z^{-j}T_j+
\frac{1}{2}<\rho,\rho>),\tag 4.1
$$
where $T_j$ are the Sugawara elements:
$$
T_j=\frac{1}{2}\sum_{a\in B}\sum_{n\in Z}:a[n]a[j-n]:,\tag 4.2
$$
 $B$ is an orthonormal basis of $\g$ with respect to $<,>$,
$a[n]=a\otimes t^n\in \hat\g$,
$:a[n]a[m]:$ equals $a[n]a[m]$ when $m>n$ and $a[m]a[n]$ otherwise.
\endproclaim

\demo{Proof} It is known (cf. \cite{FR}, Eq. (4.42))
that near the point $q=1$ the quantum currents
have the expansion
$$
L_V^{\pm}(z)=1\otimes 1+(q-q^{-1})\sum_{a\in B}J_a^{\pm}(z)\otimes
\pi_V(a)+O((q-q^{-1})^2),\tag 4.3
$$
where $J_a^{\pm}(z)$ are classical currents:
$$
J_a^{\pm}(z)=\pm\biggl(\frac{1}{2}a^0+a^{\mp}+\sum_{n>0}a[\mp n]z^{\pm
n}\biggr),\ a=a^0+a^++a^-, a^0\in {\frak h},a^{\pm}\in {\frak n^{\pm}}.\tag 4.4
$$

This implies that
$$
\gather
L_V(z)(1\otimes q^{2\rho})=1\otimes 1+(q-q^{-1})(\sum_{a\in B}J_a(z)\otimes
\pi_V(a)+1\otimes\rho)\\
-(q-q^{-1})^2\biggl(\sum_{a,b\in B}J_a^+(z)J_b^-(z)\otimes\pi_V(ab)
+Q_V^+(z)+Q_V^-(z)\biggr)+O((q-q^{-1})^3),\tag 4.5
\endgather
$$
where $J_a=J_a^+-J_a^-$, and $Q_V^{\pm}$ are quadratic terms.

Let us compute the expansion of $l_V(z)$ using (4.5). Since $\Tr|_V(a)=0$,
$a\in\g$, we have (near $q=1$):
$$
l_V(z)=\text{dim}V-(q-q^{-1})^2\biggl(C_V\sum_{a\in B}J_a^+(z)J_a^-(z)+
K_V^+(z)+K_V^-(z)\biggr),\tag 4.6
$$
where $K_V^{\pm}(z)=\Tr|_V(Q^{\pm}_V(z))$
are quadratic terms
lying in the Borel subalgebras $U(\hat\frak{b^{\mp}})$.

{}From formula (4.6) we see that the limit (4.1) exists and equals $C_V\tilde
T(z)$, where
$$
\tilde T(z)=\sum_{a\in B}J_a^+(z)J_a^-(z)+
K_V^+(z)+K_V^-(z).\tag 4.7
$$
Our purpose is to prove that $\tilde T(z)=T(z)$.
This is the same as to show that their Fourier components are the
same: $\tilde T[n]=T[n]$.

Combining (4.2), (4.6), we see that $T[n]-\tilde T[n]$ is
in $U(\hat\frak{b^{+}})$ if $n>0$, in $U(\hat\frak{b^{-}})$
if $n<0$, and in both if $n=0$. On the other hand,
both $T[n]$ and $\tilde T[n]$ are central and of degree $n$,
which implies that so is their difference. These two facts
 immediately imply
that this difference is zero for $n\ne 0$ and a constant
independent of the module $U$ if $n=0$.

To find this constant, let us assume that $U$ is a Verma module
with highest weight $\lambda$. Then $T_0=\frac{1}{2}<\l,\l+2\rho>$.
Thus we get
$$
\gather
\lim_{q\to
1}\frac{l_V^0-\text{dim}V}{(q-q^{-1})^2}=\frac{1}{2}
d^2\text{Ch}_V(x)|_{x=1}(\l+\rho,\l+\rho)=C_V<\l+\rho,\l+\rho>=
{C_V}(T_0+\frac{<\rho,\rho>}{2}),\tag 4.8
\endgather
$$
Q.E.D.
$\square$\enddemo

So far the highest weight $\l$ has been a formal parameter.
Now we specialize $\l$.

\proclaim{Corollary 4.2} Let $\g=\frak{sl}_2$. Then Lemma 3.2
holds for arbitrary special value of $\l$; Theorem 1.5 holds
for any $\l$ such that $<\l+\rho,\alpha^{\vee}>$ is not a positive
integer for any root $\alpha$ of $\g$ (positive or negative).
\endproclaim

\demo{Proof} The statement follows by deformation argument
from Theorem 4.1 and the fact that $T_i$ are algebraically independent
in any Verma module.
\enddemo

{\bf Remark. } Probably, Corollary 4.2 is true for any $\g$.
However, the method we used to prove Theorem 1.5
does not seem to be powerful enough to show it: for example,
Lemma 3.1 is false for $\l=-\rho$ even for $\g=\frak{sl}_2$.
\vskip .1in

{\bf Acknowledgements. } We are grateful to our advisor Igor Frenkel
for inspiring discussions. We thank E.Frenkel, V.Ginzburg, I.Grojnowski,
K. Hasegawa, A.Kirillov Jr., H.Knight, F.Malikov, N.Reshetikhin,
V.Tarasov, and A.Varchenko
for useful discussions.

This work was partially started when we were visiting Kyoto in the
summer of 1993. We thank M.Jimbo and T.Miwa for their hospitality.

The work of the second author was supported by Alfred P.Sloan
graduate dissertation fellowship.

\end
\Refs
\widestnumber\key{AAA}

\ref\key CP\by Chari, V., and Pressley, A.\paper Fundamental
representations of Yangians and singularities of $R$-matrices\jour
Jour f\"ur die Reine und angew. Math.\vol 417\yr 1991\pages 87-128
\endref

\ref\key DF\by Ding, J. and Frenkel, I.B.\paper
Isomorphism of two realizations of Quantum affine algebra
$U_q(\widehat{\frak gl}(n))$\jour Comm. Math. Phys. \vol 156\pages
277-300\yr 1993\endref

\ref\key Dr1\by Drinfeld, V.G. \paper Quantum groups\inbook Proc. Int.
Congr. Math., Berkeley, 1986\pages 798--820\endref

\ref\key Dr2 \by Drinfeld, V.G.\paper On almost cocommutative Hopf
algebras \jour Leningrad Math.J. \vol 1\issue 2\yr 1990\pages
321--342\endref

\ref\key Dr3\by Drinfeld, V.G. \paper Hopf algebras and the
quantum Yang-Baxter equations\jour Soviet Math. Dokl.\vol 32\yr
1985\pages 254-258\endref

\ref\key Dr4\by Drinfeld, V.G. \paper A new realization of Yangians
and quantized affine algebras\jour Soviet Math. Dokl.\vol 36\yr
1988\pages 212-216\endref

\ref\key FF\by Feigin, B.L. and Frenkel, E.V.\paper
Affine Kac-Moody algebras at the critical level and Gelfand-Dikii
algebras
\jour Int. Jour. Mod. Phys. A\vol 7 \issue Suppl 1A\yr 1992\pages
197-215
\endref

\ref\key FFR\by Feigin, B.L., Frenkel, E.V., and Reshetikhin,
N.Yu.\paper Gaudin model, Bethe ansatz, and correlation functions
at the critical level\jour hep-th 9402022\yr 1994\endref

\ref\key FR\by  Frenkel, I.B., and Reshetikhin, N.Yu.\paper Quantum affine
algebras and holonomic difference equations
\jour Comm. Math. Phys.\vol
146\pages 1-60\yr 1992\endref

\ref \key FRT \by Reshetikhin, N.Yu., Takhtadzhyan, L.A. and Faddeev,
L.D.\paper Quantization of Lie groups and Lie algebras\jour Leningrad
Math. J. \vol 1\issue 1\yr 1990\pages 193--225\endref

\ref\key J \by Jimbo, M.A.\paper A q-analog of U(gl(N+1)), Hecke
algebras, and the Yang-Baxter equation\jour Lett. Math. Phys.\vol
11\yr 1986\pages 247-252\endref

\ref\key KhT\by Khoroshkin, S.M., and Tolstoy, V.N.\paper
On Drinfeld's realization of quantum affine algebras\jour
Jour. of Geom. and Phys.\yr 1993\endref

\ref\key R\by Reshetikhin, N.Yu. \paper Quasitriangle Hopf algebras
and invariants of tangles\jour Leningrad Math J. \vol 1\issue 2 \pages
491-513\yr 1990\endref

\ref\key RS\by Reshetikhin, N.Yu. and Semenov-Tian-Shansky, M.A.
\paper Central extensions of quantum current groups\jour Lett. Math.
Phys.
\vol 19\pages 133-142\yr 1990\endref

\endRefs
